\documentclass[aps,prd,nofootinbib,superscriptaddress]{revtex4}

\usepackage{graphicx}
\usepackage{natbib}
\usepackage{amsmath}  
\usepackage{amsfonts}
\usepackage{amssymb}
\usepackage{natbib} 
\usepackage{graphicx} 
\usepackage{xspace}
\usepackage{color}
\usepackage{url}
\usepackage{braket}

\usepackage{ulem}
\usepackage{color}
\def\bm{\boldmath}

\begin{document}

\title{
\vspace{-1.5cm}
 \begin{flushright}
    LFTC-24-01/84
  \end{flushright}
  \vspace{0.5cm}
$B_c^{\pm}$-$^{12}$C states and detailed study of momentum space method for
$\Upsilon$- and $\eta_b$-nucleus bound states}

\author{G.~N.~Zeminiani}
\email{guilherme.zeminiani@gmail.com}
\affiliation{Laborat\'orio de F\'isica Te\'orica e Computacional (LFTC),
Programa de P\'{o}sgradua\c{c}\~{a}o em Astrof\'{i}sica e F\'{i}sica Computacional,
Universidade Cidade de S\~ao Paulo (UNICID), 
01506-000, S\~ao Paulo, SP, Brazil}

\author{J.J. Cobos-Mart\'inez}
\email{jesus.cobos@unison.mx}
\affiliation{Departamento de F\'isica, Universidad de Sonora, Boulevard Luis Encinas J. y Rosales,
Colonia Centro, Hermosillo, Sonora
83000, M\'exico}

\author{K.~Tsushima}
\email{kazuo.tsushima@gmail.com,  kazuo.tsushima@cruzeirodosul.edu.br}
\affiliation{Laborat\'orio de F\'isica Te\'orica e Computacional (LFTC),
Programa de P\'{o}sgradua\c{c}\~{a}o em Astrof\'{i}sica e F\'{i}sica Computacional,
Universidade Cidade de S\~ao Paulo (UNICID),
01506-000, S\~ao Paulo, SP, Brazil}





\begin{abstract}
We perform a detailed study of the $\Upsilon$-, $\eta_b$-, and $B_c$-nucleus
systems in momentum space to calculate
the bound-state energies and the corresponding coordinate-space radial wave functions.
The attractive strong potentials for the
meson-nucleus systems are calculated from the Lorentz scalar mass modifications of
these mesons in nuclear matter in the local density approximation in the nucleus.
The downward shift of the meson masses may be regarded as a signature
of partial restoration of chiral symmetry
in a nuclear medium applied in the present study in an empirical sense,
because the origin of the negative mass shift in this study is not
directly related to the chiral symmetry mechanism.
Furthermore, as an initial and realistic study, the $B_c^{\pm}$-$^{12}$C
bound states are studied for the first time,
with the effects of self-consistently
calculated Coulomb potentials in $^{12}$C (when the $B_c^{\pm}$ mesons are absent).
\end{abstract}

\maketitle

\section{Introduction}

Studies of hadron properties under extreme conditions indicate that
the Lorentz scalar effective masses
of mesons are expected to decrease in a nuclear medium, as a consequence of
partial restoration of chiral
symmetry~\cite{Krein:2010vp,Tsushima:2011kh,Ko:1992tp,Asakawa:1992ht}.
This negative effective mass shift
(Lorentz scalar) of the meson can be regarded as
an attractive Lorentz scalar potential,
and when the sum of the Lorentz scalar and the Lorentz vector potentials
(total potential in a nonrelativistic sense) is sufficiently
attractive, the mesons can be bound to atomic nuclei.
In 1985, Friedman and Soff discussed deeply bound
pionic states in heavy atomic nuclei~\cite{Friedman:1984yg},
which were first observed in 1996 in the 
$^{208}$Pb($d$, $^3$He) reaction~\cite{Yamazaki:1996ch}.
That led to predictions of other possible meson-nucleus bound
states~\cite{Hayano:1998sy,Tsushima:1998qw,Tsushima:1998ru}.
In the case of heavy quarkonia (no vector potentials arise in the
lowest order), charmonium-nucleus systems were proposed in
1989~\cite{Brodsky:1989jd}, which led to various subsequent
studies~\cite{Krein:2010vp,Tsushima:2011kh,Hosaka:2016ypm,Metag:2017yuh,Krein:2017usp,
Lee:2000csl,Krein:2013rha,Klingl:1998sr,Hayashigaki:1998ey,Kumar:2010hs,Belyaev:2006vn,
Yokota:2013sfa,Peskin:1979va,Kharzeev:1995ij,Kaidalov:1992hd,Luke:1992tm,deTeramond:1997ny,
Brodsky:1997gh,Sibirtsev:2005ex,Voloshin:2007dx,TarrusCastella:2018php,Cobos-Martinez:2020ynh}
and lattice QCD calculations~\cite{Yokokawa:2006td,Kawanai:2010ev,Skerbis:2018lew}
on these states.
Recent lattice results show also a possibility of $\phi$-nucleon ($N$) bound
states~\cite{Chizzali:2022pjd}, although $\phi$ is not the heavy quarkonium.
Furthermore, by the developed Faddeev three-body approach using the lattice extracted
$\phi$-$N$ potential, a possible $\phi$-$NN$ bound state is predicted~\cite{Etminan:2024vkv}.
We have studied further the bottomonium-nucleus
systems~\cite{Krein:2010vp,Cobos-Martinez:2020ynh}
and predicted that the $\Upsilon$ and $\eta_b$ mesons can form strong nuclear
bound states with various
nuclei~\cite{Zeminiani:2020aho,Zeminiani:2021vaq,Zeminiani:2021xvw,Cobos-Martinez:2022fmt},
provided that they are produced inside a nucleus with very low
relative momenta to the nucleus.

The Okubo-Zweig-Iizuka rule dictates the suppression of
the light hadron exchange
in the heavy quarkonium-nucleus interaction, and then the interaction must occur
primarily by multigluon exchange,
namely, a QCD van der Waals type of
interaction~\cite{Brodsky:1997gh}.
Looking at different possibilities,
we consider an alternative (effective) mechanism
for the quarkonium-nucleus interaction.
We estimated the charmonia ($J/\Psi$ and $\eta_c$)
and bottomonia ($\Upsilon$ and $\eta_b$) mass modifications (Lorentz scalar) by
the enhanced in-medium self-energies via the excitations of intermediate-state
hadrons with light quarks~\cite{Cobos-Martinez:2020ynh}.

The focus of the present study is first the bottomonia,
for which the self-energies were calculated including
only the minimal loop contributions for the $\Upsilon$ and $\eta_b$ mesons,
namely, the $BB$ loop for the $\Upsilon$, and the $BB^*$ loop for the
$\eta_b$~\cite{Zeminiani:2020aho}.
The calculations are performed neglecting any possible imaginary part of the self-energies.
Second, we study the $B_c^{\pm}$-nucleus bound states for the first time, with and without
the Coulomb force.

To obtain the bound-state energies and the corresponding 
bound meson state
wave functions of the meson-nucleus systems in the coordinate space,
we solve the Klein-Gordon equation in momentum space,
where for the $\Upsilon$ case, the Proca equation is approximated and reduced to the
Klein-Gordon equation assuming the longitudinal and transverse modes are nearly
equal with low momentum.
It is very convenient to deal with the kinetic term in the Klein-Gordon equation
in momentum space, and the search for the bound-state energies is also much simpler
than performing in coordinate space.
This can be done by solving the differential wave equation as an eigenvalue equation,
and using a method for finding selected eigenvalues and eigenfunctions (eigen vectors).
This makes it easy to extend the calculation for complex Hamiltonians,
e.g., when including the meson widths and/or imaginary potentials.
In coordinate space, dealing with complex eigenvalues is tricky,
as it involves searching for a convergence point in the real
and imaginary parts in energy plane, as well as the real and imaginary parts
of the coordinate-space wave functions.

The bound-state energies calculated for $\Upsilon$-$^{12}$C,
$\Upsilon$-$^{4}$He, $\eta_b$-$^{12}$C and $\eta_b$-$^{4}$He systems reported
in Ref.~\cite{Cobos-Martinez:2022fmt}, are calculated again here and compared
with the present results,
for which we present in this study for the first time
the wave functions in coordinate space associated with each energy level of the bound state.
In this article we also report the first results of the
$B_c$-nucleus bound states, for
$B_c$-$^4$He and $B_c$-$^{12}$C with no Coulomb interaction, followed by a
realistic study of $B_c^{\pm}$-$^{12}$C including
the Coulomb potentials,
where the Coulomb potentials are calculated self-consistently in the $^{12}$C nucleus
when the $B_c^{\pm}$ mesons are absent,
within the quark-meson coupling (QMC) model treatment for finite
nuclei~\cite{Guichon:1987jp,Guichon:1995ue,Tsushima:2002cc,Tsushima:1997df,
Saito:1996sf,Saito:2005rv,Tsushima:2019wmq}.

This article is organized as follows.
In Sec.~\ref{scmss} we discuss the downward shift masses of the $\Upsilon$, $\eta_b$ and
$B_c$ mesons in nuclear matter.
In Sec.~\ref{scbst} we relate the mass shift with the attractive meson-nucleus potentials 
and discuss the method used to solve
the Klein-Gordon equation in momentum space.
The numerical procedure of the calculation is explained in 
Sec.~\ref{scnmr}, and the results are presented in Sec.~\ref{scrslt}.
An initial study of the $B_c$-$^4$He (no Coulomb potential)
and $B_c^{\pm}$-$^{12}$C bound states (with and without the Coulomb potentials)
is made in Sec.~\ref{scbc}. The summary and conclusions are given in Sec.~\ref{sccncl}.

\section{Meson mass shift in nuclear matter}
\label{scmss}

\begin{figure}[!htb]%
\centering
\includegraphics[scale=0.33]{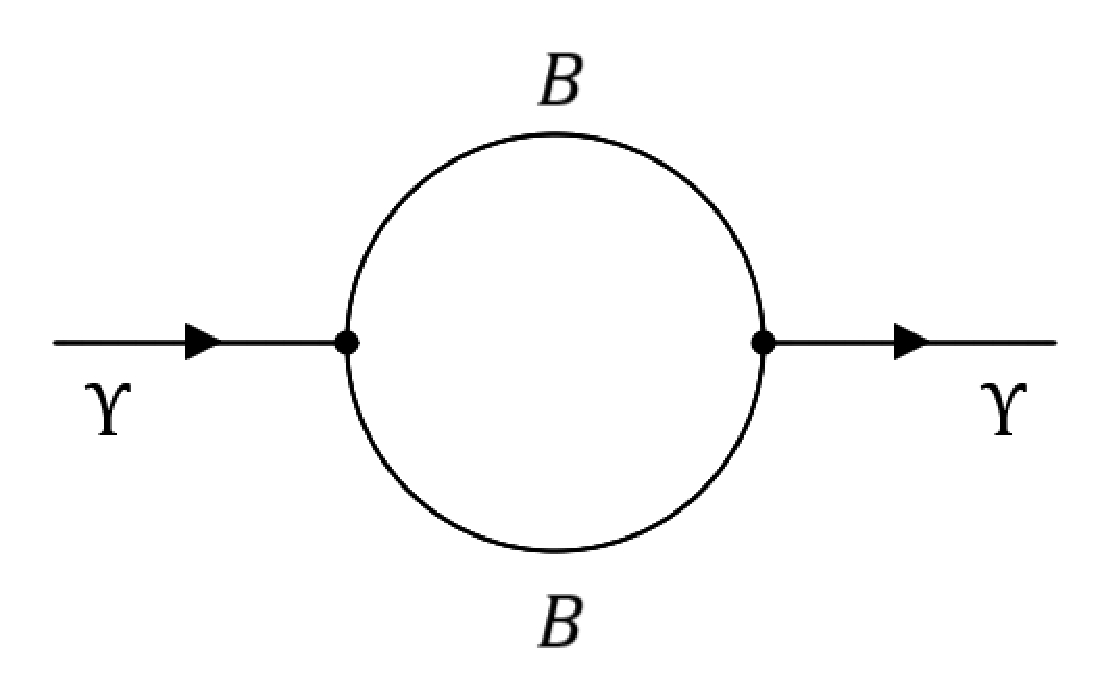}
\hspace{2ex}
\includegraphics[scale=0.33]{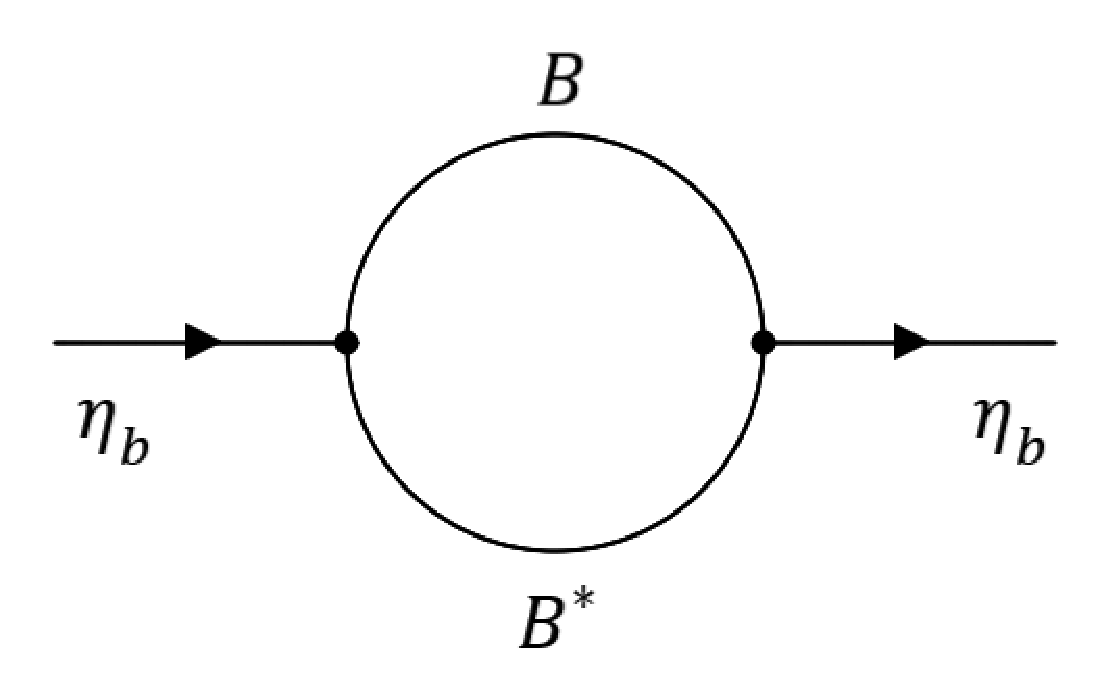}
\includegraphics[scale=0.33]{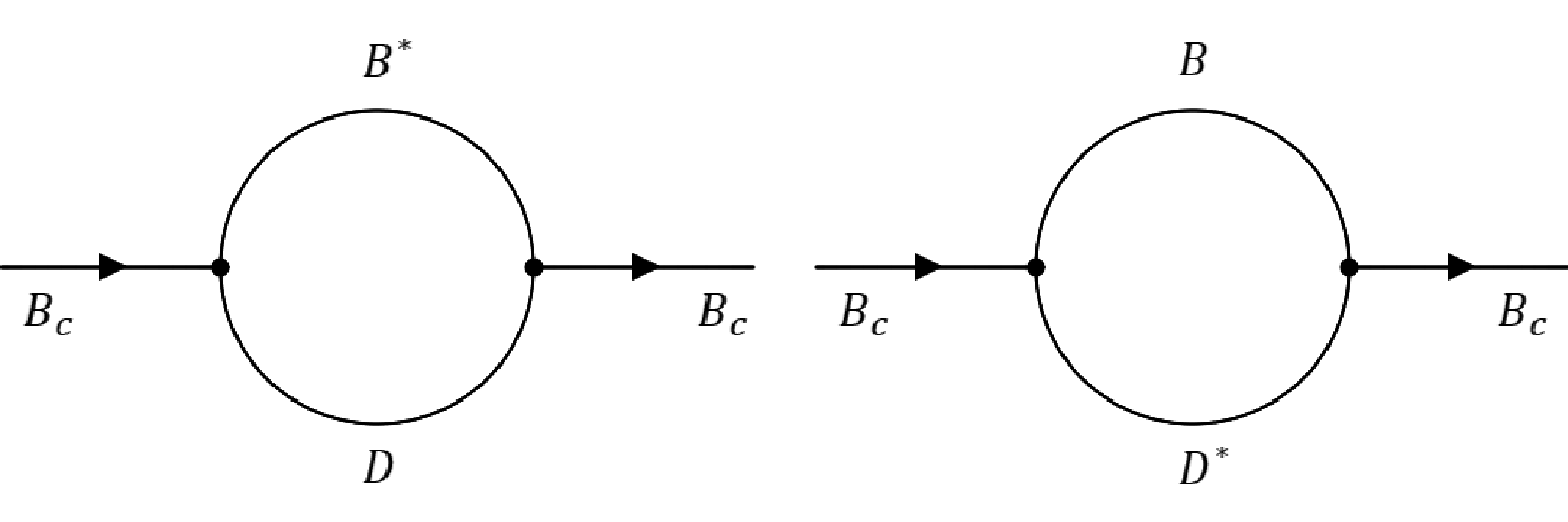}
\caption{Diagrammatic representation of the lowest order one-loop contributions
to the $\Upsilon$ (top left), $\eta_b$ (top right) and $B_c$ (bottom)
self-energies.}
\label{selfen}
\end{figure}

The mass shift of the $\Upsilon$, $\eta_b$, and $B_c$ mesons originates from
the medium modifications of the intermediate state meson loop contributions
to their self-energies, represented in Fig.~\ref{selfen}.
The self-energies are calculated based on the following 
effective Lagrangian (densities) for each vertex derived
from a flavor SU(5) symmetric effective Lagrangian (density)~\cite{Zeminiani:2020aho}

\begin{eqnarray}
{\cal L}_{\Upsilon BB} 
&=& i g_{\Upsilon BB}\Upsilon^{\mu}
\left[\overline{B} \partial_{\mu}B 
 - \left(\partial_{\mu}\overline{B}\right)B\right], \nonumber
\label{lag1}
\\
{\cal L}_{\eta_b BB^*} 
&=& i g_{\eta_b BB^*}
\left\{ (\partial^\mu \eta_b) 
\left( \overline{B^*}_\mu B - \overline{B} B^*_\mu \right)
- \eta_b 
\left[ \overline{B^*}_\mu (\partial^\mu B) - (\partial^\mu \overline{B}) B^*_\mu \right]
\right\}, \nonumber
\label{lag2}
\\
\mathcal{L}_{B_{c}B^{*}D} &=& ig_{B_{c}B^{*}D}
    [(\partial_{\mu}B^{-}_{c}){D}
    - B^{-}_{c}(\partial_{\mu}{D})] B^{* \mu} + H.c., \nonumber
\label{lag3}
\\
\mathcal{L}_{B_{c}BD^{*}} &=& ig_{B_{c}BD^{*}}
    [(\partial_{\mu}B^{+}_{c})\overline{B}
    - B_c^{+}(\partial_{\mu}\overline{B})] \overline{D^{*}}^\mu + H.c.,
\label{lag4},
\end{eqnarray}
where the following conventions are used
\begin{align*}
B&=\begin{pmatrix}
        B^{+}\\
        B^{0}
       \end{pmatrix}, & \overline{B}=\begin{pmatrix}
       B^{-} & \overline{B}^{0} \end{pmatrix},
			& &B^{*} =\begin{pmatrix}
        B^{*+}\\
        B^{*0}
       \end{pmatrix}, & &\overline{B^{*}}&=\begin{pmatrix}
       B^{*-} & \overline{B}^{*0} \end{pmatrix},\\
\end{align*}

\begin{align*}
\overline{D}&=\begin{pmatrix}
        \overline{D}^{0}\\
        D^{-}
       \end{pmatrix}, & D=\begin{pmatrix}
       D^{0} & D^{+} \end{pmatrix},  
			& &\overline{D}^{*} =\begin{pmatrix}
        \overline{D}^{*0}\\
        D^{*-}
       \end{pmatrix}, & &D^{*}&=\begin{pmatrix}
       D^{*0} & D^{*+} \end{pmatrix}.\\         
\end{align*}

The SU(5) symmetric universal coupling $g$
is fixed by
$g_{\Upsilon BB} = \frac{5g}{4\sqrt{10}} \approx 13.2$
by the $\Upsilon$ decay data $\Gamma(\Upsilon \to e^+ e^-)$
with the vector meson dominance (VMD) model~\cite{Zeminiani:2020aho},
and thus we get
\begin{equation}
g_{\eta_b BB^*} = g_{\Upsilon BB} \approx 13.2,
\hspace{3ex}
g_{B_c B^* D} = g_{B_c B D^*} = \frac{2}{\sqrt{5}}g_{\Upsilon BB} 
= \frac{g}{2\sqrt{2}}
\approx 11.9.
\end{equation}

The in-medium Lorentz scalar potential, or the mass shift value,
for the heavy mesons $h$(=$\Upsilon$,$\eta_b$, $B_c$)
is given by the difference of the
in-medium $m_{h}^{*}$ and the free space $m_{h}$ masses

\begin{equation}
 V = m_{h}^{*} - m_{h},
\end{equation}
with the free-space physical mass being self-consistently reproduced by
\begin{equation}
m^{2}_{h} = \left(m^{0}_{h}\right)^{2} - 
|\Sigma_{h} (k^{2}=m^{2}_{h})|,
\label{phymass}
\end{equation}
where $m^{0}_{h}$ is the bare mass and $\Sigma_{h}$ is the total self-energy. 
The in-medium mass $m_{h}^{*}$ is calculated likewise, 
with the same bare mass
value $m^0_{h}$ determined in free space,
and the in-medium masses of the
$B$, $B^*$, $D$, and $D^*$ mesons are calculated using the
QMC model~\cite{Guichon:1987jp,Guichon:1995ue,Saito:1996sf}.
Neglecting any possible effects of the
widths, the self-energies in medium are given below for each meson

\begin{eqnarray}
\Sigma^{BB}_{\Upsilon}(m^{*}_{\Upsilon}) &=&
(- \frac{g^{2}_{\Upsilon BB}}{3\pi^{2}})
\int_{0}^{\infty} d|\textbf{k}| 
    |\textbf{k}|^{2} I_{\Upsilon}^{BB}(|\textbf{k}|)
    F_{\Upsilon BB} (\textbf{k}^2) \nonumber
\\
\Sigma^{BB^{*}}_{\eta_{b}}(m^{*}_{\eta_{b}}) &=& 
\frac{8 g_{\eta_{b} B B^*}^{2}}{\pi^{2}}\int_{0}^{\infty} 
d|\textbf{k}| 
    |\textbf{k}|^{2} I_{\eta_{b}}^{BB^{*}}(|\textbf{k}|)
    F_{\eta_{b} BB^{*}} (\textbf{k}^2) \nonumber
\\
\Sigma^{B^{*}D}_{B_{c}}(m^{*}_{B_{c}}) &=& 
\frac {-4g^{2}_{B_{c}B^{*}D}}{\pi^{2}} \int d|\textbf{k}| 
    |\textbf{k}|^{2} I_{B_c}^{B^{*}D}(|\textbf{k}|)
    F_{B_c B^* D} (\textbf{k}^2),
\end{eqnarray}
where

\begin{eqnarray}
I_{\Upsilon}^{BB}(|\textbf{k}|) &=&
\frac{1}{{\omega^*_{B}}}
\left(\frac{|\textbf{k}|^{2}}{{\omega^{*2}_{B}
- m^{*2}_{\Upsilon}}/4}\right), \nonumber
\\
I_{\eta_{b}}^{BB^{*}}(|\textbf{k}|) &=&
\left. \frac{m_{\eta_{b}}^{*2}(-1+(k^0)^2/m_{B^{*}}^{*2})}
{(k^{0}+\omega^*_{B^{*}})(k^{0}-\omega^*_{B^*})
(k^{0}- m_{\eta_{b}}^* -\omega^*_{B})}\right|_{k^{0}=m_{\eta_{b}}^* - \omega^*_B}
\nonumber \\
&&\hspace{5ex} + \left. \frac{m_{\eta_{b}}^{*2}(-1+(k^0)^2/m_{B^{*}}^{*2})}
{(k^{0}-\omega^*_{B^{*}})(k^{0}-m_{\eta_{b}}^*+\omega^*_{B})
(k^{0}-m_{\eta_{b}}^* -\omega^*_{B})}\right|_{k^{0}=-\omega^*_{B^*}}, \nonumber
\\
I_{B_c}^{B^{*}D} (|\textbf{k}|) &=&
\left. \frac{m^{*2}_{B_c} \left(-1 + (k^0)^2 / m^{*2}_{B^*} \right)}
    {(k^0 - \omega^*_{B^*}) (k^0 - m^{*}_{B_c} + \omega^*_D)
    (k^0 - m^{*}_{B_c} -\omega^*_D)}
    \right|_{k^0 = -\omega^*_{B^*} }
    \nonumber \\
    && \left. \hspace{5ex} + \frac{m^{*2}_{B_c} \left( -1 + (k^0)^2 / m^{*2}_{B^*} \right) }
    {(k^0 + \omega^*_{B^*}) (k^0 - \omega^*_{B^*})
    (k^0 - m^{*}_{B_c} -\omega^*_D)}
    \right|_{k^{0} = m^{*}_{B_c} - \omega^*_D},
\end{eqnarray}
and $\omega^*_{B,B^*,D,D^*}=(\textbf{k}^{2}+m_{B,B^*,D,D^*}^{*2})^{1/2}$.
[A similar calculation has been performed for
$\Sigma^{BD^{*}}_{B_{c}}(m^{*}_{B_{c}})$, and will be reported in the future.]
To regularize the divergence in the loop integral
we introduce into the integrand a regularizing function given
by the product of the vertex form factors.
We use the following form factors for the vertex in-medium as
(now the in-medium meson masses enter, instead of free meson masses):
$F_{B_c B^* D} (\textbf{k}^2) = u_{B_cB^*}(\textbf{k}^2)u_{B_c D}(\textbf{k}^2)$,
where $u_{B_c B^*} = \left( \frac{\Lambda^2_{B^*} + m^{*2}_{B_c}}
{\Lambda^2_{B^*} + 4\omega^{*2}_{B^*} (\textbf{k}^2)} \right)^2$ and
$u_{B_c D} = \left( \frac{\Lambda^2_{D} + m^{*2}_{B_c} }
{\Lambda^2_{D} + 4\omega^{*2}_{D} (\textbf{k}^2)} \right)^2$
with $\Lambda_{B^*}$ and $\Lambda_{D}$ being the cutoff masses associated 
with the $B^*$ and $D$
mesons, respectively.
Similar form factors are used for the other vertices. We use a common
value for the cutoff parameter
$\Lambda \equiv \Lambda_{B,B^*,D,D^*}$,
ranging
$\Lambda$ = 2000, 3000, 4000, 5000, and 6000 MeV
and see the ambiguities by the values on the final results.

Solving Eq.~(\ref{phymass})
with the self-energy calculated with the medium-modified
intermediate-state meson masses, we estimate the
nuclear density dependent mass shift, presented in
Fig.~\ref{deltam} as a function of the nuclear matter density
($\rho_B/\rho_0$) for the different
values of the cutoff mass $\Lambda$.

\begin{figure}[htb]%
\vspace{6ex}
\includegraphics[width=5.5cm]{1_BB.eps}
\hspace{2ex}
\includegraphics[width=5.5cm]{etab_BBs.eps}
\hspace{2ex}
\includegraphics[width=5.5cm]{Bc_totalpot.eps}
\caption{In-medium mass shift of $\Upsilon$ (left), $\eta_b$ (center) and $B_c$ (right)
mesons versus baryon density ($\rho_B/\rho_0$) for five different
values of the cutoff mass $\Lambda$.}
\label{deltam}
\end{figure}

\section{Nuclear bound states}
\label{scbst}

\begin{figure}[htb!]
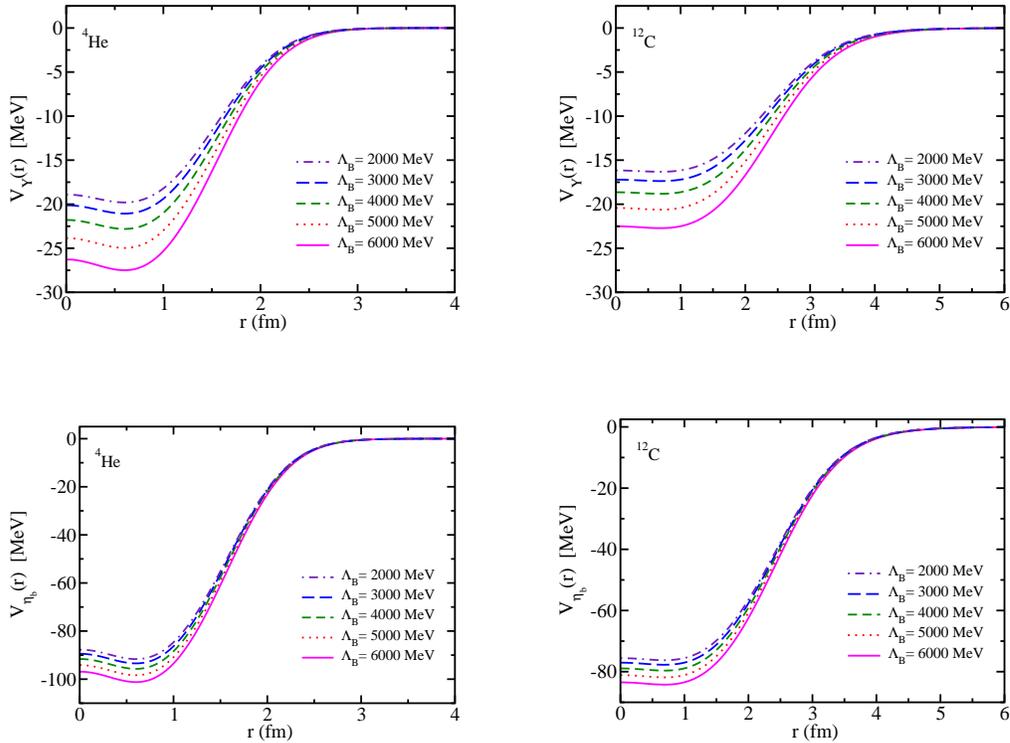
%
\vspace{8ex}
\hspace{-16ex}
\includegraphics[width=6.0cm]{Upsipot_He4.eps}
\hspace{8ex}
\includegraphics[width=6.0cm]{Upsipot_C12.eps}
\\
\vspace{8ex}
\hspace{-16ex}
\includegraphics[width=6.0cm]{etabpot_He4.eps}
\hspace{8ex}
\includegraphics[width=6.0cm]{etabpot_C12.eps}
\caption{Nuclear (Lorentz scalar) potentials for the
$\Upsilon$-$^4$He (top-left),
$\Upsilon$-$^{12}$C (top-right),
$\eta_b$-$^4$He (bottom-left)
and $\eta_b$-$^{12}$C (bottom-right)
systems for different values of the cutoff parameter
$\Lambda$
.}%
\label{npots}%
\end{figure}

First, we study the bottomonium-nucleus bound states.
The coordinate-space nuclear potentials for the systems
$\Upsilon$-$^{4}$He, $\Upsilon$-$^{12}$C, $\eta_b$-$^{4}$He and $\eta_b$-$^{12}$C
are calculated using a local density approximation.
For a bottomonium $h$(=$\Upsilon$,$\eta_b$) and a nucleus $A$, the potential
is given by the equation
\begin{equation}
    V_{h-A} (r=|\vec{r}|)
    = \Delta m_{h} \left( \rho^{A}_{B} (r) \right)
    \equiv m_h^*(r) - m_h,
\end{equation}
where $r$ is the distance from the center of the nucleus,
$\Delta m_{h}$ is the value of the meson mass shift (Lorentz scalar),
$m^*_h$ being the in-medium effective $\Upsilon$ or $\eta_b$ mass in nuclear matter,
and $\rho^{A}_{B} (r)$
is the nuclear density distribution in the nucleus $A$.
For the $^{12}$C nucleus, it was calculated by the QMC model~\cite{Saito:1996sf}, and for the
$^4$He nucleus we use the parameterized nuclear density 
from Ref.~\cite{Saito:1997ae}. The potentials are shown in Fig.~\ref{npots}.

The meson-nucleus bound states are studied first
in momentum space by solving the Klein-Gordon (K.G.) equation.
While the K.G. equation is solved for the pseudoscalar $\eta_b$ meson, we also solve the K.G.
equation for the spin-1 $\Upsilon$ meson.
This is based on the approximation (assuming the case) that the $\Upsilon$-nucleus relative
momentum is nearly zero.
In this situation, the transverse and longitudinal components in the Proca equation are
expected to be very similar,
which allows us to reduce it to solving a single-component K.G. equation, as practiced in
Refs.~\cite{Zeminiani:2021vaq,Zeminiani:2021xvw,Cobos-Martinez:2022fmt}.
The $\eta_b$- and $\Upsilon$-nuclear potentials 
were previously calculated
in coordinate space in Ref.~\cite{Cobos-Martinez:2022fmt},
in which they were transformed into the
momentum-space by one particular method
for that study (``Woods-Saxon Fourier transform'' method to be explained later).
For the purpose of the present study, we will transform
the coordinate-space potentials using another method.
A comparison of the results obtained by the different methods is presented in the
Appendix~\ref{ApxA}.

We solve the K.G. equation in momentum space of the form
\begin{equation}
\label{kgeq}
\left[ \vec{P}^{2} + \left( m + V_{N} \right)^{2} \right] \psi
= \varepsilon^{2}\psi,
\end{equation}
where $\vec{P}$ is the "relative momentum'' of the meson and the nucleus,
$\psi$ is the momentum-space wave function of the system,
$m$ is the reduced mass of the meson-nucleus system
($m = m_h m_A/(m_h + m_A)$, $m_h$ is the meson mass and $m_A$ the nucleus mass)
and $V_N$ is the Lorentz scalar nuclear potential in momentum space, where 
$V_N(r) \equiv V_{h-A} (r)$.
The K.G. equation will be solved for each value of the angular momentum
$\ell$ of the meson-nucleus system to calculate the bound-state energies
and to get the momentum-space eigen wave function
corresponding to each bound-state energy level of the system.

The partial wave solutions for the bound-state energies and the corresponding
wave functions of the K.G equation require the decomposition of
the meson-nucleus potential into a given angular momentum $\ell$ wave.
Generally, a straightforward way to obtain the $\ell$-wave decomposition
of the meson-nucleus potential
$V_N(r)$ is to apply the double spherical Bessel transform,
\begin{equation}
\label{bsstr}
V_{N \ell}(P,P') = \frac{2}{\pi} \int^{\infty}_{0} dr r^{2}
j_{\ell}(Pr) j_{\ell}(P'r) V_{N}(r),
\end{equation}
where $j_\ell$ ($\ell$ = 0, 1, 2, $\cdots$) are the spherical Bessel functions.
(The reason why the potential has two arguments $P$ and $P'$
will become clearer in the next section.)

Since this transformation contains spherical Bessel functions, the oscillatory behavior of
these functions must be tamed properly in order to construct the wave function in momentum space,
as well as when the momentum-space wave function is transformed to the coordinate
space for each partial wave $\ell$ solution.
This is achieved by tuning the $r$ and $P$ grids to construct suitable discrete sets.
More explanations on this will be given in a later section.

\section{Numerical Procedure}
\label{scnmr}

The partial wave expansion of Eq.~(\ref{kgeq}) is given by
\begin{eqnarray}
\label{lwvkg}
\varepsilon^{2}_{n\ell} \psi_{n\ell}(P) &=&
(P^2 + m^2)\psi_{n\ell}(P)
+ 2m \int^{\infty}_0 dP' P^{'2} V_{N \ell} (P, P') \psi_{n\ell}(P')
\nonumber \\
&+& \int^{\infty}_0 dP' P^{'2} \left[
\int^{\infty}_0 dP'' P''^{2} V_{N \ell} (P, P'') V_{N \ell} (P'',P')
\right] \psi_{n\ell}(P').
\end{eqnarray}

The discretized Eq.~(\ref{lwvkg}) can be written in matrix form as
\begin{equation}
\label{mtxeq}
\sum^{NP}_{j=1} \hat{H}_{ij} \hat{\psi}_j
= \varepsilon^{2}_{n\ell} \hat{\psi}_i,
\end{equation}
where $\hat{\psi}_i = \psi_{n\ell}(P_i)$. The matrix elements of the Hamiltonian are
\begin{equation}
\label{hamdisc}
\hat{H}_{ij} = (P^2_i + m^2)\delta_{ij} + 2m P^2_j w_j V_{N \ell}(P_i,P_j)
+ \left( \sum^{NP}_{k=1} P^2_k w_k
V_{N \ell}(P_i,P_k) V_{N \ell}(P_k,P_j) \right) P^2_j w_j,
\end{equation}
where $NP$ is the number of points for the momentum grid,
and $P_i$ and $w_j$ are respectively the $i$th momentum point and $j$th weight.

Equation~(\ref{mtxeq}) can be solved as an eigenvalue equation using iterative methods,
such as the inverse iteration~\cite{Heddle:1985dh}.
It is a method for finding selected eigenvalues
and eigenfunctions, that has the advantage of only requiring the inversion
of an $NP \times NP$ matrix.
The equation to be solved is of the type
\begin{equation}
\label{invp}
H \ket{\psi_{n}} = \varepsilon_{n} \ket{\psi_{n}}.
\end{equation}

We begin with some guess value
$\tau_n$, and constructing the operator
\begin{equation}
B_{n} = (H - \tau_{n})^{-1},
\end{equation}
that contains the Hamiltonian of Eq.~(\ref{invp}).
Next, we construct a state $\ket{\chi^{0}_{n}}$
that can be expressed using 
the complete set of eigenfunctions $\ket{\psi_{n}}$
\begin{equation}
\ket{\chi^{0}_{n}} = \sum_{n'} C_{n'} \ket{\psi_{n'}},
\end{equation}
which is multiplied $N$ times by the operator $B_{n}$,
resulting in
\begin{equation}
\ket{\chi^{N}_{n}} = (B_{n})^{N} \ket{\chi^{0}_{n}} = \sum_{n'} C_{n'}
(\varepsilon_{n'} - \tau_{n})^{-N} \ket{\psi_{n'}}.
\end{equation}

For nondegenerate eigenstates and for a proper guess $\tau_n$, $n' = n$ will rapidly take
over the sum. Then, for a large enough N
\begin{eqnarray}
\ket{\chi^{N}_{n}} &\approx& C_{n} (\varepsilon_{n} 
- \tau_{n})^{-N} \ket{\psi_{n}}, \nonumber \\
\ket{\chi^{N+1}_{n}} &\approx& (\varepsilon_{n} - \tau_{n})^{-1} \ket{\chi^{N}_{n}},
\end{eqnarray}
which can be written in a momentum-space calculation as
\begin{eqnarray}
\chi^{N}_{n}(P) &\approx& C_{n} (\varepsilon_{n} - \tau_{n})^{-N} \psi_{n}(P), \nonumber \\
\chi^{N+1}_{n}(P) &\approx& (\varepsilon_{n} - \tau_{n})^{-1} \chi^{N}_{n}(P).
\end{eqnarray}

To facilitate the iterative numerical calculation, 
$\chi^{N}_{n}(P)$ is divided by its element with maximum modulus, 
$\Tilde{\chi}^{N}_{n}(P) = \chi^{N}_{n}(P)/|\chi^{N}_{n}(P)|_{\text{max}}$,
and $\chi^{N+1}_{n}(P)$ is redefined as
\begin{equation}
\chi^{N+1}_{n}(P) \equiv B_n \Tilde{\chi}^{N}_{n}(P),
\end{equation}
and for sufficiently large $N$,
\begin{equation}
\label{egsts}
|\chi^{N}_{n}(P)|_{\text{max}} = \frac{1}{\varepsilon_{n} - \tau_{n}},
\quad
\Tilde{\chi}^{N}_{n}(P) = \frac{\psi_n (P)}{|\psi_n (P)|_{\text{max}}}.
\end{equation}

The eigenvalue is then given by
\begin{equation}
  \varepsilon_{n} =  \tau_{n}
  + \frac{1}{|\chi^{N}_{n}(P)|_{\text{max}}}.
\end{equation}

In the same way, the eigenvector $\psi_n (P)$ is determined by Eq.~(\ref{egsts}).
The obtained eigenvalue and eigenvector are
the system's eigenenergy and momentum-space wave function, respectively,
for each $n$ and $\ell$.

\subsection{Selection of gridpoints}
\label{sbscgrd}

There are two sets of gridpoints used to perform the numerical calculations:
The $r$ and the $P$ gridpoints. These points must be properly chosen to be able to perform
correctly
the Gaussian integrations in Eqs.~(\ref{bsstr}) 
and~(\ref{hamdisc}).
In particular, the choice of grid in momentum space is not trivial
compared to that in coordinate space, and one needs some care.

For the double spherical Bessel transform,
one is recommended to properly limit the maximum number of points,
so that the spherical Bessel functions, depending on both $r$ and $P$ grids, do not cause many oscillations.
This is necessary to obtain the stable and nonexcessively oscillating wave
functions (more explanations later).
Furthermore, the $r$ grid should
contain the proper and sufficient points in the region
where the potential is relevant.
For example, the potentials presented in Fig.~\ref{npots} are
relevant in the region from $r$ = 0 fm up to $r$ = 2 or 3 fm.
This means that most of the $r$ gridpoints should be in this
region close to the center of the nucleus,
leaving fewer points to the region outside of this.
A good choice for the momentum gridpoints in this integration is found
to be $P \approx 1/r$.

When constructing the Hamiltonian, however, the $P$ distribution of points
is not necessarily the same as when performing the spherical Bessel transform.
This is because of the kinetic term in the Hamiltonian.
When transforming the potential to the momentum space, one should consider primarily
the points where the potential is relevant, whereas in constructing the Hamiltonian of the
system, Eq.~(\ref{hamdisc}), the points in the momentum corresponding to the
regions in coordinate space further away from the center of nucleus must also be included. 
We use the points
$P_i = \tan \left[\frac{1}{4} \pi (1 + x_i)\right]$ following Ref.~\cite{Kwon:1978tz},
where $x_i$ are the Gaussian points defined
in the region $x_i=[-1,1]$.

\section{Results}
\label{scrslt}

After solving numerically Eq.~(\ref{mtxeq}), we can obtain
the bound-state energies of the meson-nucleus system, that are given by 
$E_{n\ell} = \varepsilon_{n\ell} - m$. The results are presented in Table~\ref{tblbse}
for the central value of the cutoff parameter $\Lambda = 4000$ MeV.
A comparison for different values of $\Lambda$ is presented in the Appendix~\ref{ApxA}
so that one can have ideas on the $\Lambda$-value dependence.
We have obtained the energy levels by varying the eigenvalue
initial guess, $\tau_n$, for fixed $\ell$.
To get each eigenvalue, a different initial guess value is chosen, and this is repeated
until all the desired energy levels are obtained.
Each converged eigenvalue is then confirmed by
analysing the number of nodes in the corresponding coordinate-space wave function.
(Detailed explanations are given in next subsection.)

\begin{table}[htb!]
\caption{\label{tblbse} $^{4}_{\Upsilon}\text{He}$, $^{12}_{\Upsilon}\text{C}$,
$^{4}_{\eta_b}\text{He}$ and $^{12}_{\eta_b}\text{C}$
bound state energies
for the central value of the cutoff parameter $\Lambda = 4000$ MeV.
}
\begin{center}
\begin{tabular}{ll|r}
  \hline \hline
  & \multicolumn{2}{c}{$E_{n\ell}$ (MeV)} \\
\hline
& $n\ell$ & $\Lambda = 4000$ MeV\\
\hline
$^{4}_{\Upsilon}\text{He}$
& 1s & -6.25 \\
\hline
$^{12}_{\Upsilon}\text{C}$
& 1s & -15.26 \\
& 1p &-9.57 \\
\hline
\hline
$^{4}_{\eta_b}\text{He}$
& 1s & -71.59 \\
& 1p & -41.50 \\
& 1d & -39.56 \\
& 2s & -30.09 \\
\hline
$^{12}_{\eta_b}\text{C}$
& 1s & -66.93 \\
& 1p & -55.13 \\
& 1d & -48.50 \\
& 2s & -36.30 \\
& 1f & -28.09 \\
& 2p & -20.67 \\
\hline
\end{tabular}
\end{center}
\end{table}

\subsection{Coordinate-space wave functions}
\label{sbscrwf}

From the eigenvectors that come out as solutions of Eq.~(\ref{mtxeq}),
we can use a (single) spherical Bessel transform to obtain
the coordinate-space wave functions of the corresponding energy levels of the system:
\begin{equation}
    \psi_{n \ell}(r=|\vec{r}|) = \frac{1}{2\pi} \int^{\infty}_{0}
dP\, P^2\, j_{\ell}(Pr)\, \psi_{n \ell}(P),
\end{equation}
where the normalization is performed
in $\psi_{n \ell}(P)$ according to the convention
(recall that the wave functions are real in the present study),
\begin{equation}
\int \psi_{n \ell} (P) \psi_{n \ell} (P) P^2 dP = 1.
\end{equation}

As commented already, this procedure enables us to confirm that the obtained eigen energies are
indeed corresponding to the correct eigen wave functions.

The gridpoints used in this integration may better be 
the same as those used to obtain the partial wave
decomposition of the nuclear potential, namely the same as those when performing
the double spherical Bessel transform.
Otherwise the resultant wave function would
be often unstable, with exceeding oscillations.
We have observed that, the more the grid points differ from the corresponding points
used for the partial wave decomposition, the more oscillations in the wave function are.
For this reason, when one performs the spherical Bessel transform of the potential,
one needs to chose properly the gridpoints that result in a smooth wave function.

The results for the $\Upsilon$-$^4$He, $\Upsilon$-$^{12}$C,
$\eta_b$-$^4$He, and $\eta_b$-$^{12}$C wave functions, for the central value of
the cutoff parameter $\Lambda = 4000$ MeV are presented in Fig.~\ref{psiryetc}.

\begin{figure}[htb!]%
\vspace{8ex}
\hspace{-16ex}
\includegraphics[width=6.0cm]{Psir_Y_He4_4000_1s.eps}
\hspace{8ex}
\includegraphics[width=6.0cm]{Psir_Y_C12_4000_total.eps}
\\
\vspace{8ex}
\hspace{-16ex}
\includegraphics[width=6.0cm]{Psir_etab_He4_4000_total.eps}
\hspace{8ex}
\includegraphics[width=6.0cm]{Psir_etab_C12_4000_total.eps}
\caption{Coordinate-space wave functions for the $\Upsilon$-$^4$He (top left), $\Upsilon$-$^{12}$C (top right),
$\eta_b$-$^4$He (bottom left) and $\eta_b$-$^{12}$C (bottom right) systems
for the central value of the cutoff parameter $\Lambda = 4000$ MeV.}%
\label{psiryetc}%
\end{figure}

\section{\bf\bm Initial study of the $B_c^{\pm}$-$^{4}$He and $B_c^{\pm}$-$^{12}$C
bound states with and without the Coulomb potentials}
\label{scbc}

We now proceed to an initial study of the $B_c^{\pm}$-$^{4}$He
(without the Coulomb potentials) and $B_c^{\pm}$-$^{12}$C
bound states (with and without the Coulomb potentials).
The mass shift of $B_c$ meson in symmetric nuclear matter was calculated in
Ref.~\cite{Zeminiani:2023gqc}, and we obtain the $B_c$-nucleus strong potential
using a local density approximation.
For a realistic calculation,
the Coulomb potentials should enter for $B_c^\pm$-nucleus potentials,
but they are absent in the $\Upsilon$ and $\eta_b$ cases.
First, we include only
the strong interaction part of the
$B_c^{\pm}$-$^{4}$He and $B_c^{\pm}$-$^{12}$C potentials,
without the Coulomb potentials.
Second, including the Coulomb potentials, we study the
$B_c^{\pm}$-$^{12}$C bound states, and focus on the role of the Coulomb potentials.
This is because the $^{12}$C nucleus is constructed self-consistently
by the QMC model (relativistic, quark-based nuclear shell model),
but this is not done for the $^{4}$He nucleus.

The strong interaction potentials for $B_c$ are presented
in Fig.~\ref{bcnuc}.
These are calculated in the same manner as those for the $\Upsilon$ and $\eta_b$ cases.
We solve the K.G. equation and obtain the bound-state
energies for the central value of the cutoff parameter $\Lambda = 4000$ MeV,
which we present in
Table~\ref{tblbccent}.
The corresponding wave functions are shown in Fig.~\ref{psirbcstr}.
\begin{figure}[htb!]%
\vspace{6ex}
\centering
\includegraphics[width=6.0cm]{Bc_nucl_pot_He4.eps}
\hspace{8ex}
\includegraphics[width=6.0cm]{Bc_nucl_pot_C12.eps}
\caption{Nuclear potentials for the $B_c$-$^4$He and $B_c$-$^{12}$C 
systems for different values of the cutoff parameter $\Lambda$.}
\label{bcnuc}
\end{figure}

\begin{table}
\caption{\label{tblbccent} $^{4}_{B_c}\text{He}$ and $^{12}_{B_c}\text{C}$ bound-state
  energies without the Coulomb potentials 
  for the central value of the cutoff parameter $\Lambda = 4000$ MeV.}
\begin{center}
\begin{tabular}{ll|r}
  \hline \hline
  & \multicolumn{2}{c}{$E_{n\ell}$ (MeV)} \\
\hline
& $n\ell$ & $\Lambda = 4000$ MeV\\
\hline
$^{4}_{B_c}\text{He}$
& 1s & -76.67 \\
& 1p & -42.91 \\
\hline
$^{12}_{B_c}\text{C}$
& 1s & -76.06 \\
& 1p & -52.88 \\
\hline
\end{tabular}
\end{center}
\end{table}

\begin{figure}[htb!]%
\vspace{6ex}
\centering
\includegraphics[width=6.0cm]{Psir_Bc_He4_4000_1s1p.eps}
\hspace{8ex}
\includegraphics[width=6.0cm]{Psir_Bc_C12_4000_1s1p.eps}
\caption{Coordinate-space wave functions for the $B_c^{\pm}$-$^{4}$He (left) and $B_c^{\pm}$-$^{12}
$C (right) systems without the Coulomb potentials
for the central value of the cutoff parameter $\Lambda = 4000$ MeV.}
\label{psirbcstr}
\end{figure}

We now study the more realistic case for the $B_c^{\pm}$-$^{12}$C,
where we include the Coulomb interaction.
Including the Coulomb potential $V_c$ 
in Eq.~(\ref{kgeq}), it becomes
\begin{equation}
\left[ \vec{P}^2 + \left( m + V_N \right)^2 \right] \psi = \left( \varepsilon - V_c \right)^2 \psi,
\label{kgecoul}
\end{equation}
and the Hamiltonian for the numerical calculation in Eq.~(\ref{hamdisc}) now becomes
\begin{eqnarray}
\label{hamcdisc}
\hat{H}_{ij} &=& (P^2_i + m^2)\delta_{ij} + 2m P^2_j w_j V_{N \ell}(P_i,P_j)
+ \left( \sum^{NP}_{k=1} P^2_k w_k
V_{N \ell}(P_i,P_k) V_{N \ell}(P_k,P_j) \right) P^2_j w_j \nonumber \\
&+& 2\varepsilon_{n\ell} P^2_j w_j V_{c \ell}(P_i,P_j)
- \left( \sum^{NP}_{k=1} P^2_k w_k
V_{c \ell}(P_i,P_k) V_{c \ell}(P_k,P_j) \right) P^2_j w_j.
\end{eqnarray}

The Coulomb potentials in the $^{12}$C nucleus
(nucleon density distribution and Coulomb mean field)
is obtained within
the QMC model~\cite{Saito:1996sf} self-consistently.
We show the Coulomb potentials side-by-side with the nuclear potentials
for this nucleus in Fig.~\ref{CoulPot}. Since the mean-field approximation of the QMC model
is not good for a light nucleus such as the $^4$He, we do not
investigate the $B_c^{\pm}$-$^{4}$He
system with the Coulomb potentials in this initial study.
In the near future, we plan to include the Coulomb potentials for the
$B_c^{\pm}$-$^{4}$He systems with a different approach, although the effects
are expected to be small.

\begin{figure}[htb]%
\vspace{6ex}
\centering
\includegraphics[width=7.5cm]{Bc_nucl_Coul_pot_C12.eps}
\caption{Attractive and repulsive Coulomb potentials, together with the nuclear potentials for the
$^{12}_{B^{\pm}_c}\text{C}$ systems.}
\label{CoulPot}
\end{figure}

The treatment of the Coulomb potential in momentum space is known to be
difficult due to the logarithmic singularity that appears in its partial wave decomposition,
requiring the use of some regularization technique,
such as the Lande subtraction method~\cite{Heddle:1985dh,Kwan:1978zh}.
In the present calculation, however, the treatment is much more simple.
The reason is that we can calculate the $B_c^{\pm}$-$^{12}$C
Coulomb potentials in coordinate space using the
self-consistent mean Coulomb field
in the $^{12}$C nucleus by neglecting the
feedback of the Coulomb force from $B_c^{\pm}$ (which should give very small effects).
Thus, we simply get the direct partial wave decomposition of the Coulomb
potentials in momentum space according to the method already explained
in Sec.~\ref{scbst}.

We solve Eq.~(\ref{kgecoul}) using the direct double spherical Bessel transform for the
following cases:
(i) nuclear and attractive Coulomb potentials  ($B_c^-$-$^{12}$C),
(ii) nuclear and repulsive Coulomb potentials  ($B_c^+$-$^{12}$C),
(iii) only nuclear potentials,
(iv) only Coulomb potentials,
and (v) the cases of the term $2\varepsilon V_c \to 0$ in the right-hand side of
Eq.~(\ref{kgecoul}),
on the right-hand side of the expansion, 
$(\varepsilon-V_c)^2 = \varepsilon^2 + V_c^2 - 2 \varepsilon V_c$.
The results are presented in Table~\ref{tblcoul4} (see also Table~\ref{tblcoul} in
Appendix~\ref{ApxA}).
We make this comparison to see the interference effect
between the dominant strong nuclear and Coulomb potentials, which arises
naturally for the relativistic K.G. equation case but not for the lowest-order
Schr\"{o}dinger equation.
Solving Eq.~(\ref{kgecoul}) for $n =1$, $\ell = 0$ and the Coulomb potential
alone, we get
energies, $E_{n\ell} = E_{10} \approx \pm 1.4$~MeV for $\pm |V_c|$,
with a small asymmetry between the cases of the attractive and repulsive Coulomb
potentials, that arises from the $V^2_c$ term 
and diminishes the energy $E$
in both cases. Naively one could expect that the bound-state energies
obtained including the nuclear and attractive
(or repulsive) Coulomb potentials to be around 1.4 MeV deeper (or shallower)
than the case of the nuclear potential alone, however, in fact we obtain
more than 3 to 4 MeV difference for the 1s state. This happens because the nuclear potential
dominates $\varepsilon$ in the term $2\varepsilon V_c$ that appears when we include the Coulomb
potential.
The ''interference'' between the nuclear and Coulomb potentials is
responsible for this larger difference.
By making the term $2\varepsilon V_c \to 0$, we nearly recover the energies
found for the case of only nuclear interaction,
aside from a small difference due to the $V^2_c$ term.
This shows that when including the Coulomb interaction,
the nuclear potential dominates $\varepsilon$, as one can expect.
Although the Coulomb effect is small, such behavior described above
cannot be seen in a naive treatment using the Schr\"{o}dinger equation without
including up to the $V_c^2$ term.

\begin{table}
\caption{\label{tblcoul4} $^{12}_{B^{\pm}_c}\text{C}$ bound-state
energies with the Coulomb potentials
for the central value of the cutoff parameter $\Lambda = 4000$ MeV.}
\begin{center}
\begin{tabular}{ll|c}
  \hline \hline
  & \multicolumn{2}{c}{$E_{n\ell}$ (MeV)} \\
  \hline
& \multicolumn{2}{c}{$B_c$-$^{12}$C (Strong only)} \\
\hline
& $n\ell$ & $\Lambda_{B}= 4000$ MeV\\
\hline
$^{12}_{B_c}\text{C}$
& 1s & -76.06 \\
& 1p & -52.88 \\
\hline
& \multicolumn{2}{c}{$B^{-}_c$-$^{12}$C} \\
\hline
& $n\ell$ & $\Lambda_{B}= 4000$ \\
\hline
$^{12}_{B^{-}_c}\text{C}$
& 1s & -80.63 \\
& 1p & -57.53 \\
\hline
& \multicolumn{2}{c}{$B_c^-$-$^{12}$C ($2EV_c \rightarrow 0$)} \\
\hline
& $n\ell$ & $\Lambda_{B}= 4000$ \\
\hline
$^{12}_{B^{-}_c}\text{C}$
& 1s & -76.58 \\
& 1p & -54.01 \\
\hline
& \multicolumn{2}{c}{$B^{+}_c$-$^{12}$C} \\
\hline
& $n\ell$ & $\Lambda_{B}= 4000$ \\
\hline
$^{12}_{B^{+}_c}\text{C}$
& 1s & -72.53 \\
& 1p & -50.49 \\
\hline
& \multicolumn{2}{c}{$B^{\pm}_c$-$^{12}$C (Coulomb only)} \\
\hline
& $n\ell$ & $B_c^-$-$^{12}$C \\
\hline
$^{12}_{B^{\pm}_c}\text{C}$
& 1s & -1.43 \\
\hline
& $n\ell$ & $B_c^-$-$^{12}$C ($2EV_c \rightarrow 0$) \\
\hline
$^{12}_{B^{\pm}_c}\text{C}$
& 1s & -0.01 \\
\hline
& $n\ell$ & $B_c^+$-$^{12}$C \\
\hline
$^{12}_{B^{\pm}_c}\text{C}$
& 1s & +1.46 \\
\hline
\end{tabular}
\end{center}
\end{table}

The coordinate-space wave functions for the 1s and 1p states for
the $B^{\pm}_c$-$^{12}$C system for the central value of 
the cutoff parameter $\Lambda = 4000$ MeV are presented in Fig.~\ref{psirbcstrpm}.

\begin{figure}[htb!]%
\vspace{6ex}
\centering
\includegraphics[width=6.0cm]{Psir_Bc_AttCoul_C12_4000_1s1p.eps}
\hspace{8ex}
\includegraphics[width=6.0cm]{Psir_Bc_RepCoul_C12_4000_1s1p.eps}
\caption{Coordinate-space wave functions for 1s and 1p states of the $B^{-}_c$-$^{12}$C (left) and
$B^{+}_c$-$^{12}$C (right) systems for the central value of the cutoff parameter
$\Lambda = 4000$ MeV.
\label{psirbcstrpm}
}
\end{figure}

\section{Summary and Conclusion}
\label{sccncl}

By solving the Klein-Gordon equation in momentum space, we have calculated the single-particle
energies and the corresponding coordinate-space wave functions
for the quarkonium-nucleus
systems, for the $\Upsilon$ and $\eta_b$ mesons and the $^4$He and $^{12}$C nuclei.
The meson-nucleus potentials have been calculated from the mass shift amount
of the mesons in nuclear matter using a local density approximation without the effects
of the meson widths. 
The results depend on the cutoff parameter $\Lambda$ values,
introduced in the regularization of the meson self-energy calculations.

We also have calculated for the first time the
$B_c$-$^4$He and $B_c$-$^{12}$C bound-state
energies and coordinate-space wave functions,
with only the strong interaction
potentials first, and then,
for the $B_c^{\pm}$-$^{12}$C system, we have studied including
the realistic Coulomb potentials, and analyzed the ''interference'' effect
between the nuclear and Coulomb potentials.

In the future, more realistic studies can be made for
the $\Upsilon$-, $\eta_b$- and $B_c^{\pm}$-nucleus systems for various nuclei,
including the possible imaginary
part of the potentials (widths) for the $\Upsilon$-, $\eta_b$-
and $B_c^{\pm}$-nucleus bound states.
In particular, the realistic Coulomb potentials will be included
for the $B_c^{\pm}$- and $B_c^{* \pm}$-nucleus systems.


\begin{acknowledgments}
The authors acknowledge the support and warm hospitality of
Asia Pacific Center for Theoretical Physics (APCTP) during the Workshop (APCTP PROGRAMS 2023)
''Origin of Matter and Masses in the Universe: Hadrons in free space, dense nuclear medium,
and compact stars,'' where important discussions and development were achieved
on the topic.
The authors also thank the Origin of Matter and Evolution of Galaxies Institute at Soongsil
University for the supports in many aspects during the collaboration visit in Korea.
The authors thank A.~W.~Thomas for his useful comments on the coordinate space wave functions at
the initial stage of the study and fruitful discussions on this topic.
G.N.Z.~was supported by the Coordena\c{c}\~ao de Aperfei\c{c}oamento de Pessoal
de N\'ivel Superior-Brazil (CAPES).
JJCM acknowledges financial support from the University of Sonora under Grant No. USO315009105.
K.T.~was supported by Conselho Nacional de Desenvolvimento
Cient\'{i}fico e Tecnol\'ogico (CNPq, Brazil), Processes No.~313063/2018-4,
No.~426150/2018-0, and No.~304199/2022-2,
and FAPESP Process No.~2019/00763-0 and No.~2023/073-3-6 (G.N.Z. and K.T.).
The work of G.N.Z and K.T. was in the projects of
Instituto Nacional de Ci\^{e}ncia e
Tecnologia - Nuclear Physics and Applications
(INCT-FNA), Brazil, Process No.~464898/2014-5.
\end{acknowledgments}




\newpage


\appendix
\section[\appendixname~\thesection]{Results comparison}
\label{ApxA}

The treatment in momentum space requires the nuclear potentials to be transformed from
coordinate space to momentum space, and then decomposed into partial waves.
For this purpose we compare three different methods, namely,
(i) the spherical Bessel transform of the numerically obtained original potential in
coordinate space, (ii) the partial wave decomposition of the Fourier transform of the
Woods-Saxon
approximated form for the original potential, and
(iii) the spherical Bessel transform of the Woods-Saxon approximation of
the numerically obtained original potential.
The spherical Bessel transform of the numerically obtained original potential was already explained in Sec.~\ref{scbst}.

In order to use conveniently the approximated form for the calculated
quarkonium-nucleus potentials shown in Fig.~\ref{npots},
we adopt the Woods-Saxon potential form which is commonly used to approximately
parameterize nuclear potentials,
\begin{equation}
\label{wseq}
    V_{WS} (r=|\vec{r}|) = -\frac{V_0}{1 + e^{(r-R)/a}},
\end{equation}
where $V_0 (> 0)$ is the potential depth, $R$ is the nuclear radius and
$a$ is the surface thickness of the nucleus.
These parameters will be determined by the fit for each nucleus
for a given numerically obtained original potential.

The use of the Woods-Saxon form parameterized potential enables
us to work with an analytical form of $V_N (r)$. This is done by interpolating
the numerical data to fit to the Woods-Saxon shape.
The obtained parameters for each potential are presented in
Tables~\ref{tblws} and~\ref{tblwsbc}.
The advantage of using the Woods-Saxon potential form is that,
one does not need to be provided with the obtained original numerical potential data,
but just may know the Woods-Saxon potential parameters.
\begin{table}[htb!]%
\caption{\label{tblws} Parameters of the fitted Woods-Saxon form potentials for
the $V_{\Upsilon - ^4\text{He}}$,
$V_{\Upsilon - ^{12}\text{C}}$,
$V_{\eta_b - ^4\text{He}}$, and $V_{\eta_b - ^{12}\text{C}}$ potentials and
different cutoff values.}
\begin{center}
\begin{tabular}{ll|r|r|r}
  \hline \hline
  & & \multicolumn{3}{c}{Woods-Saxon potential parameter values} \\
\hline
& & $\Lambda_{B}=2000$ MeV& $\Lambda_{B}= 4000$ MeV&
$\Lambda_{B}= 6000$ MeV\\
\hline
$V_{\Upsilon - ^4\text{He}}$
& $V_0$ (MeV)& 19.83 & 22.84 & 27.54 \\
& $R$ (fm)& 1.6199 & 1.6215 & 1.6237 \\
& $a$ (fm)& 0.2832 & 0.2833 & 0.2835 \\
\hline
$V_{\Upsilon - ^{12}\text{C}}$
& $V_0$ (MeV)& 16.66 & 19.21 & 23.20 \\
& $R$ (fm)& 2.4638 & 2.4662 & 2.4692 \\
& $a$ (fm)& 0.4805 & 0.4806 & 0.4808 \\
\hline \hline
$V_{\eta_b - ^4\text{He}}$
& $V_0$ (MeV)& 91.84 & 95.91 & 101.42 \\
& $R$ (fm)& 1.6351 & 1.6352 & 1.6353 \\
& $a$ (fm)& 0.28445 & 0.28446 & 0.28447 \\
\hline
$V_{\eta_b - ^{12}\text{C}}$
& $V_0$ (MeV)& 77.82 & 81.28 & 85.96 \\
& $R$ (fm)& 2.4854 & 2.4856 & 2.4858 \\
& $a$ (fm)& 0.48181 & 0.48182 & 0.48183 \\
\hline \hline
\end{tabular}
\end{center}
\end{table}
%

\begin{table}[htb!]%
\caption{\label{tblwsbc} Parameters of the fitted Woods-Saxon
shape for the $V_{B_c - ^4\text{He}}$ and
$V_{B_c - ^{12}\text{C}}$ potentials and different cutoff values.}
\begin{center}
\begin{tabular}{ll|r|r|r}
  \hline \hline
  & & \multicolumn{3}{c}{Woods-Saxon potential parameter values} \\
\hline
& & $\Lambda_{B}=2000$ MeV& $\Lambda_{B}= 4000$ MeV&
$\Lambda_{B}= 6000$ MeV\\
\hline
$V_{B_c - ^4\text{He}}$
& $V_0$ (MeV)& 112.12 & 114.61 & 125.45 \\
& $R$ (fm)& 1.6308 & 1.6311 & 1.6307 \\
& $a$ (fm)& 0.28427 & 0.2843 & 0.28426 \\
\hline
$V_{B_c - ^{12}\text{C}}$
& $V_0$ (MeV)& 94.77 & 96.90 & 106.04 \\
& $R$ (fm)& 2.4796 & 2.4801 & 2.4795 \\
& $a$ (fm)& 0.48167 & 0.4817 & 0.48166 \\
\hline \hline
\end{tabular}
\end{center}
\end{table}
%

A comparison between the numerically obtained original
potentials and the corresponding fitted Woods-Saxon form potentials is made
in Fig.~\ref{wsap} for
$V_{\Upsilon-^4{\rm He}}$ and $V_{{\Upsilon}-^{12}{\rm C}}$
for the case of $\Lambda = 2000$ MeV.
One can easily observe that the fitted Woods-Saxon form potentials are good
approximations of the original potentials, and the differences in the calculated bound
state energies using the parametrized potentials are expected to be small
(indeed, this will be demonstrated later).
\begin{figure}[htb!]%
\vspace{6ex}
\centering
\includegraphics[width=6.0cm]{He4_2000_ws_data_comp.eps}
\hspace{8ex}
\includegraphics[width=6.0cm]{C12_2000_ws_data_comp.eps}
\caption{
Comparison between the originally obtained
$\Upsilon$-nucleus potentials (solid line)
and the fitted Woods-Saxon potentials (long-dashed line)
for $^4$He (left) and $^{12}$C (right) for the cutoff $\Lambda=2000$ MeV.
\label{wsap}
}
\end{figure}

The Fourier transform of the Woods-Saxon form potential
of Eq.~(\ref{wseq}) can be carried out analytically
using complex analysis. After the integration in the complex plane, we end up
with the following equation
\begin{eqnarray}
\label{wsfr}
   V_{WS} (\vec{P},\vec{P}') &=& -\frac{V_0}{2\pi^2} \frac{a^2}{|\vec{P}-\vec{P}'|} 
\Biggl\{ \frac{2\pi e^{-\pi \rho}}{\left(1 - e^{-2\pi \rho}\right)^2}
\biggr[ \pi \left(1 - e^{-2\pi \rho}\right) \sin(\alpha \rho)
-\alpha \left(1 - e^{-2\pi \rho}\right) \cos(\alpha \rho) \biggr]
\nonumber \\
&-&2 \sum^{n_{max}}_{n=0} (-1)^{n} \gamma^{n}
\frac{n \rho}{(\rho^2 + n^2)} \Biggl\},
\end{eqnarray}
with $\rho = (|\vec{P}-\vec{P}'|)a$, $\alpha = R/a$ and $\gamma = e^{-\alpha}$.
The value of $n_{max}$ does not affect significantly the
final result, since the first few terms in the sum already
gives nearly a converged result, and thus $n_{max}$ = 4 or 5 is sufficient.

The partial wave $\ell$-projected potentials in momentum space are then calculated by
\begin{equation}
\label{frdcmp}
    V_{N \ell} (P,P') = 2 \pi \int^{1}_{-1} d\cos\theta ~
P_{\ell} (\cos \theta) ~ V_{WS}(\vec{P},\vec{P}'),
\end{equation}
where $P_\ell$ are the Legendre polynomials of $\ell$th order and $\theta$ is the angle between
$\vec{P}$ and $\vec{P}'$.
After the Fourier transform is performed analytically,
the partial wave decomposition is made numerically.

The results obtained by the different methods of partial wave decomposition and
different values of the cutoff values are presented in Tables~\ref{tblups} to~\ref{tblcoul}.
The wave functions are presented in Figs.~\ref{psiryhe} to~\ref{coulwfn}.

\begin{table}[htb!]
\caption{\label{tblups} $^{4}_{\Upsilon}\text{He}$ and $^{12}_{\Upsilon}\text{C}$
bound-state energies, obtained by the Direct Bessel
  [Eq.~(\ref{bsstr})], Woods-Saxon Fourier 
  [Eqs.~(\ref{wsfr}) and~(\ref{frdcmp})] and 
  Woods-Saxon Bessel [Eqs.~(\ref{wseq}) and~(\ref{bsstr})] transform methods.
The $\Lambda_B$ values are in MeV. }
\begin{center}
\begin{tabular}{ll|r|r|r}
  \hline \hline
  & & \multicolumn{3}{c}{Bound state energies (MeV)} \\
  \hline
& & \multicolumn{3}{c}{Direct Bessel transform} \\
\hline
& $n\ell$ & $\Lambda_{B}=2000$ & $\Lambda_{B}= 4000$ &
$\Lambda_{B}= 6000$ \\
\hline
$^{4}_{\Upsilon}\text{He}$
& 1s & -5.93 & -6.25 & -6.56 \\
\hline
$^{12}_{\Upsilon}\text{C}$
& 1s & -13.22 & -15.26 & -18.41 \\
& 1p & -8.30 & -9.57 & -11.51 \\
\hline
& & \multicolumn{3}{c}{Woods-Saxon Fourier transform} \\
\hline
& $n\ell$ & $\Lambda_{B}=2000$ & $\Lambda_{B}= 4000$ &
$\Lambda_{B}= 6000$ \\
\hline
$^{4}_{\Upsilon}\text{He}$
& 1s & -5.48 & -7.4 & -10.6 \\
\hline
$^{12}_{\Upsilon}\text{C}$
& 1s & -10.51 & -12.67 & -16.1 \\
& 1p & -5.95 & -7.79 & -10.78 \\
\hline
& & \multicolumn{3}{c}{Woods-Saxon Bessel transform} \\
\hline
& $n\ell$ & $\Lambda_{B}=2000$ & $\Lambda_{B}= 4000$ &
$\Lambda_{B}= 6000$ \\
\hline
$^{4}_{\Upsilon}\text{He}$
& 1s & -6.35 & -6.75 & -7.14 \\
\hline
$^{12}_{\Upsilon}\text{C}$
& 1s & -13.18 & -15.22 & -18.37 \\
& 1p & -8.18 & -9.43 & -11.33 \\
\hline
\end{tabular}
\end{center}
\end{table}

\begin{table}[htb!]
\caption{\label{tbleta} $^{4}_{\eta_b}\text{He}$ and $^{12}_{\eta_b}\text{C}$ bound-state
  energies, obtained by the Direct Bessel
  [Eq.~(\ref{bsstr})], Woods-Saxon Fourier 
  [Eqs.~(\ref{wsfr}) and~(\ref{frdcmp})] and 
  Woods-Saxon Bessel [Eqs.~(\ref{wseq}) and~(\ref{bsstr})] transform methods.
The $\Lambda_B$ values are in MeV.}
\begin{center}
\begin{tabular}{ll|r|r|r}
  \hline \hline
  & & \multicolumn{3}{c}{Bound state energies (MeV)} \\
  \hline
& & \multicolumn{3}{c}{Direct Bessel transform} \\
\hline
& $n\ell$ & $\Lambda_{B}=2000$ & $\Lambda_{B}= 4000$ &
$\Lambda_{B}= 6000$ \\
\hline
$^{4}_{\eta_b}\text{He}$
& 1s & -68.71 & -71.59 & -75.44 \\
& 1p & -39.97 & -41.50 & -43.54 \\
& 1d & -37.73 & -39.56 & -42.03 \\
& 2s & -29.14 & -30.09 & -31.38 \\
\hline
$^{12}_{\eta_b}\text{C}$
& 1s & -63.70 & -66.93 & -70.27 \\
& 1p & -53.17 & -55.13 & -59.38 \\
& 1d & -46.47 & -48.50 & -51.17 \\
& 2s & -34.53 & -36.30 & -39.43 \\
& 1f & -26.56 & -28.09 & -29.97 \\
& 2p & -18.86 & -20.67 & -23.15 \\
\hline
& & \multicolumn{3}{c}{Woods-Saxon Fourier transform} \\
\hline
& $n\ell$ & $\Lambda_{B}=2000$ & $\Lambda_{B}= 4000$ &
$\Lambda_{B}= 6000$ \\
\hline
$^{4}_{\eta_b}\text{He}$
& 1s & -63.1 & -66.7 & -71.5 \\
& 1p & -40.6 & -43.7 & -48.0 \\
& 1d & -17.2 & -19.7 & -23.2 \\
& 2s & -15.6 & -17.9 & -21.1 \\
\hline
$^{12}_{\eta_b}\text{C}$
& 1s & -65.8 & -69.0 & -73.4 \\
& 1p & -57.0 & -60.1 & -64.3 \\
& 1d & -47.5 & -50.4 & -54.4 \\
& 2s & -46.3 & -49.1 & -53.0 \\
& 1f & -37.5 & -40.2 & -43.9 \\
& 2p & -36.0 & -38.6 & -42.2 \\
\hline
& & \multicolumn{3}{c}{Woods-Saxon Bessel transform} \\
\hline
& $n\ell$ & $\Lambda_{B}=2000$ & $\Lambda_{B}= 4000$ &
$\Lambda_{B}= 6000$ \\
\hline
$^{4}_{\eta_b}\text{He}$
& 1s & -67.79 & -70.65 & -74.48 \\
& 1p & -40.42 & -41.95 & -43.99 \\
& 1d & -36.46 & -38.23 & -40.64 \\
& 2s & -28.67 & -29.60 & -30.85 \\
\hline
$^{12}_{\eta_b}\text{C}$
& 1s & -63.41 & -66.05 & -69.59 \\
& 1p & -52.90 & -55.44 & -58.87 \\
& 1d & -46.34 & -48.35 & -51.05 \\
& 2s & -34.12 & -36.41 & -39.54 \\
& 1f & -26.90 & -28.29 & -30.18 \\
& 2p & -18.72 & -20.53 & -23.02 \\
\hline
\end{tabular}
\end{center}
\end{table}

\begin{table}[htb!]
\caption{\label{tblbc} $^{4}_{B_c}\text{He}$ and $^{12}_{B_c}\text{C}$ bound-state
  energies without the Coulomb potentials, obtained by the Direct Bessel
  [Eq.~(\ref{bsstr})], Woods-Saxon Fourier 
  [Eqs.~(\ref{wsfr}) and~(\ref{frdcmp})] and 
  Woods-Saxon Bessel [Eqs.~(\ref{wseq}) and~(\ref{bsstr})] transform methods.
  The $\Lambda_B$ values are in MeV.}
\begin{center}
\begin{tabular}{ll|r|r|r}
  \hline \hline
  & & \multicolumn{3}{c}{Bound state energies (MeV)} \\
  \hline
& & \multicolumn{3}{c}{Direct Bessel transform} \\
\hline
& $n\ell$ & $\Lambda_{B}=2000$ & $\Lambda_{B}= 4000$ &
$\Lambda_{B}= 6000$ \\
\hline
$^{4}_{B_c}\text{He}$
& 1s & -75.14 & -76.67 & -83.14 \\
& 1p & -42.06 & -42.91 & -46.43 \\
\hline
$^{12}_{B_c}\text{C}$
& 1s & -74.55 & -76.06 & -82.43 \\
& 1p & -51.53 & -52.88 & -58.64 \\
\hline
& & \multicolumn{3}{c}{Woods-Saxon Fourier transform} \\
\hline
& $n\ell$ & $\Lambda_{B}=2000$ & $\Lambda_{B}= 4000$ &
$\Lambda_{B}= 6000$ \\
\hline
$^{4}_{B_c}\text{He}$
& 1s & -78.04 & -80.22 & -89.68 \\
& 1p & -51.42 & -53.34 & -61.68 \\
\hline
$^{12}_{B_c}\text{C}$
& 1s & -79.67 & -81.65 & -90.14 \\
& 1p & -68.62 & -70.50 & -78.57 \\
\hline
& & \multicolumn{3}{c}{Woods-Saxon Bessel transform} \\
\hline
& $n\ell$ & $\Lambda_{B}=2000$ & $\Lambda_{B}= 4000$ &
$\Lambda_{B}= 6000$ \\
\hline
$^{4}_{B_c}\text{He}$
& 1s & -74.10 & -75.53 & -82.09 \\
& 1p & -42.43 & -43.26 & -46.75 \\
\hline
$^{12}_{B_c}\text{C}$
& 1s & -74.31 & -75.83 & -82.21 \\
& 1p & -51.79 & -53.19 & -58.93 \\
\hline
\end{tabular}
\end{center}
\end{table}

\begin{table}[htb!]
\caption{\label{tblcoul} $^{12}_{B^{\pm}_c}\text{C}$ bound-state
energies with the Coulomb potentials, obtained by the Direct Bessel [Eq.~(\ref{bsstr})]
transform method.
The $\Lambda_B$ values are in MeV.}
\begin{center}
\begin{tabular}{ll|c|c|c}
  \hline \hline
  & & \multicolumn{3}{c}{Bound state energies (MeV)} \\
  \hline
& & \multicolumn{3}{c}{$B_c$-$^{12}$C (Strong only)} \\
\hline
& $n\ell$ & $\Lambda_{B}=2000$ & $\Lambda_{B}= 4000$ &
$\Lambda_{B}= 6000$ \\
\hline
$^{12}_{B_c}\text{C}$
& 1s & -74.55 & -76.06 & -82.43 \\
& 1p & -51.53 & -52.88 & -58.64 \\
\hline
& & \multicolumn{3}{c}{$B^{-}_c$-$^{12}$C} \\
\hline
& $n\ell$ & $\Lambda_{B}=2000$ & $\Lambda_{B}= 4000$ &
$\Lambda_{B}= 6000$ \\
\hline
$^{12}_{B^{-}_c}\text{C}$
& 1s & -79.12 & -80.63 & -87.03 \\
& 1p & -56.15 & -57.53 & -63.38 \\
\hline
& & \multicolumn{3}{c}{$B_c^-$-$^{12}$C ($2EV_c \rightarrow 0$)} \\
\hline
& $n\ell$ & $\Lambda_{B}=2000$ & $\Lambda_{B}= 4000$ &
$\Lambda_{B}= 6000$ \\
\hline
$^{12}_{B^{-}_c}\text{C}$
& 1s & -75.06 & -76.58 & -82.98 \\
& 1p & -52.63 & -54.01 & -59.85 \\
\hline
& & \multicolumn{3}{c}{$B^{+}_c$-$^{12}$C} \\
\hline
& $n\ell$ & $\Lambda_{B}=2000$ & $\Lambda_{B}= 4000$ &
$\Lambda_{B}= 6000$ \\
\hline
$^{12}_{B^{+}_c}\text{C}$
& 1s & -71.01 & -72.53 & -78.94 \\
& 1p & -49.11 & -50.49 & -56.32 \\
\hline
& & \multicolumn{3}{c}{$B^{\pm}_c$-$^{12}$C (Coulomb only)} \\
\hline
& $n\ell$ & $B_c^-$-$^{12}$C & $B_c^-$-$^{12}$C ($2EV_c \rightarrow 0$)
& $B_c^+$-$^{12}$C  \\
\hline
$^{12}_{B^{\pm}_c}\text{C}$
& 1s & -1.43 & -0.01 & +1.46 \\
\hline
\end{tabular}
\end{center}
\end{table}

\begin{figure}[htb!]%
\vspace{6ex}
\includegraphics[width=5.5cm]{Psir_Y_He4_2000_1s.eps}
\hspace{2ex}
\includegraphics[width=5.5cm]{Psir_Y_He4_4000_1s.eps}
\hspace{2ex}
\includegraphics[width=5.5cm]{Psir_Y_He4_6000_1s.eps}
\caption{Coordinate-space 1s state wave functions of the $\Upsilon$-$^4$He system
for different values of cutoff $\Lambda$ obtained by the direct Bessel transform.}
\label{psiryhe}
\end{figure}

\begin{figure}[htb!]%
\vspace{6ex}
\includegraphics[width=5.5cm]{Psir_Y_C12_2000_total.eps}
\hspace{2ex}
\includegraphics[width=5.5cm]{Psir_Y_C12_4000_total.eps}
\hspace{2ex}
\includegraphics[width=5.5cm]{Psir_Y_C12_6000_total.eps}
\caption{Coordinate-space 1s and 1p state wave functions of
the $\Upsilon$-$^{12}$C system for different values of cutoff $\Lambda$ obtained by the direct
Bessel transform.}
\label{psiryc}
\end{figure}

\begin{figure}[htb!]%
\vspace{6ex}
\includegraphics[width=5.5cm]{Psir_etab_He4_2000_total.eps}
\hspace{2ex}
\includegraphics[width=5.5cm]{Psir_etab_He4_4000_total.eps}
\hspace{2ex}
\includegraphics[width=5.5cm]{Psir_etab_He4_6000_total.eps}
\caption{Coordinate-space wave functions for the 1s to 2s states of the $\eta_b$-$^4$He system
for different values of cutoff $\Lambda$ obtained by the direct Bessel transform.}
\label{psirethe}
\end{figure}

\begin{figure}[htb!]%
\vspace{6ex}
\includegraphics[width=5.5cm]{Psir_etab_C12_2000_total.eps}
\hspace{2ex}
\includegraphics[width=5.5cm]{Psir_etab_C12_4000_total.eps}
\hspace{2ex}
\includegraphics[width=5.5cm]{Psir_etab_C12_6000_total.eps}
\caption{Coordinate-space wave functions for the 1s to 2p states of the $\eta_b$-$^{12}$C system
for different values of cutoff $\Lambda$ obtained by the direct Bessel transform.}
\label{psiretc}
\end{figure}

\begin{figure}[htb!]%
\vspace{6ex}
\includegraphics[width=5.5cm]{Psir_WS_Y_He4_2000_1s.eps}
\hspace{2ex}
\includegraphics[width=5.5cm]{Psir_WS_Y_He4_4000_1s.eps}
\hspace{2ex}
\includegraphics[width=5.5cm]{Psir_WS_Y_He4_6000_1s.eps}
\caption{Coordinate-space 1s state wave functions of the $\Upsilon$-$^{4}$He system
for different values of cutoff $\Lambda$ obtained by the Fourier transform of
the fitted Woods-Saxon form potential.}
\label{psirwsyhe}
\end{figure}

\begin{figure}[htb!]%
\vspace{6ex}
\includegraphics[width=5.5cm]{Psir_WS_Y_C12_2000_total.eps}
\hspace{2ex}
\includegraphics[width=5.5cm]{Psir_WS_Y_C12_4000_total.eps}
\hspace{2ex}
\includegraphics[width=5.5cm]{Psir_WS_Y_C12_6000_total.eps}
\caption{Coordinate-space 1s and 1p state wave functions of the
$\Upsilon$-$^{12}$C system for different values of cutoff $\Lambda$ obtained by the Fourier
transform of the fitted Woods-Saxon form potential.}
\label{psirwsyc}
\end{figure}

\begin{figure}[htb!]%
\vspace{6ex}
\includegraphics[width=5.5cm]{Psir_WS_etab_He4_2000_total.eps}
\hspace{2ex}
\includegraphics[width=5.5cm]{Psir_WS_etab_He4_4000_total.eps}
\hspace{2ex}
\includegraphics[width=5.5cm]{Psir_WS_etab_He4_6000_total.eps}
\caption{Coordinate-space wave functions for the 1s to 2s states of the $\eta_b$-$^4$He system for
different values of cutoff $\Lambda$ obtained by the Fourier transform of
the fitted Woods-Saxon form potential.}
\label{psirwsethe}
\end{figure}

\begin{figure}[htb!]%
\vspace{6ex}
\includegraphics[width=5.5cm]{Psir_WS_etab_C12_2000_total.eps}
\hspace{2ex}
\includegraphics[width=5.5cm]{Psir_WS_etab_C12_4000_total.eps}
\hspace{2ex}
\includegraphics[width=5.5cm]{Psir_WS_etab_C12_6000_total.eps}
\caption{Coordinate-space wave functions for the 1s to 2p states of the $\eta_b$-$^{12}$C system
for different values of cutoff $\Lambda$ obtained by the Fourier transform of
the fitted Woods-Saxon form potential.}
\label{psirwsetc}
\end{figure}

\begin{figure}[htb!]%
\vspace{6ex}
\includegraphics[width=5.5cm]{Psir_BWS_Y_He4_2000_1s.eps}
\hspace{2ex}
\includegraphics[width=5.5cm]{Psir_BWS_Y_He4_4000_1s.eps}
\hspace{2ex}
\includegraphics[width=5.5cm]{Psir_BWS_Y_He4_6000_1s.eps}
\caption{Coordinate-space 1s state wave functions of the $\Upsilon$-$^4$He system
for different values of cutoff $\Lambda$ obtained by the spherical Bessel transform of
the fitted Woods-Saxon form potential.}
\label{psirbwsyhe}
\end{figure}

\begin{figure}[htb!]%
\vspace{6ex}
\includegraphics[width=5.5cm]{Psir_BWS_Y_C12_2000_total.eps}
\hspace{2ex}
\includegraphics[width=5.5cm]{Psir_BWS_Y_C12_4000_total.eps}
\hspace{2ex}
\includegraphics[width=5.5cm]{Psir_BWS_Y_C12_6000_total.eps}
\caption{Coordinate-space 1s and 1p state wave functions of the
$\Upsilon$-$^{12}$C system
for different values of cutoff $\Lambda$ obtained by the spherical Bessel transform of the
fitted Woods-Saxon form potential.}
\label{psirbwsyc}
\end{figure}

\begin{figure}[htb!]%
\vspace{6ex}
\includegraphics[width=5.5cm]{Psir_BWS_etab_He4_2000_total.eps}
\hspace{2ex}
\includegraphics[width=5.5cm]{Psir_BWS_etab_He4_4000_total.eps}
\hspace{2ex}
\includegraphics[width=5.5cm]{Psir_BWS_etab_He4_6000_total.eps}
\caption{Coordinate-space wave functions for the 1s to 2s states of the $\eta_b$-$^4$He system for
different values of cutoff $\Lambda$ obtained by the spherical Bessel transform of the
fitted Woods-Saxon form potential.}
\label{psirbwsethe}
\end{figure}

\begin{figure}[htb!]%
\vspace{6ex}
\includegraphics[width=5.5cm]{Psir_BWS_etab_C12_2000_total.eps}
\hspace{2ex}
\includegraphics[width=5.5cm]{Psir_BWS_etab_C12_4000_total.eps}
\hspace{2ex}
\includegraphics[width=5.5cm]{Psir_BWS_etab_C12_6000_total.eps}
\caption{Coordinate-space wave functions for the 1s to 2p states of the $\eta_b$-$^{12}$C system
for different values of cutoff $\Lambda$ obtained the spherical Bessel transform of the
fitted Woods-Saxon form potential.}
\label{psirbwsetc}
\end{figure}

\begin{figure}[htb!]%
\vspace{6ex}
\includegraphics[width=5.5cm]{Psir_Bc_He4_2000_1s1p.eps}
\hspace{2ex}
\includegraphics[width=5.5cm]{Psir_Bc_He4_4000_1s1p.eps}
\hspace{2ex}
\includegraphics[width=5.5cm]{Psir_Bc_He4_6000_1s1p.eps}
\caption{Wave functions for the 1s and 1p states of the $B_c$-$^{4}$He system without the Coulomb
potentials for different values of $\Lambda$ calculated using the direct Bessel transform.}
\label{psirbche}
\end{figure}

\begin{figure}[htb!]%
\vspace{6ex}
\includegraphics[width=5.5cm]{Psir_Bc_C12_2000_1s1p.eps}
\hspace{2ex}
\includegraphics[width=5.5cm]{Psir_Bc_C12_4000_1s1p.eps}
\hspace{2ex}
\includegraphics[width=5.5cm]{Psir_Bc_C12_6000_1s1p.eps}
\caption{Coordinate-space wave functions for the 1s and 1p states of the $B_c$-$^{12}$C system
without the Coulomb potentials for
different values of $\Lambda$ calculated using the direct Bessel transform.}
\label{psirbcc}
\end{figure}

\begin{figure}[htb!]%
\vspace{6ex}
\includegraphics[width=5.5cm]{Psir_WS_Bc_He4_2000_1s1p.eps}
\hspace{2ex}
\includegraphics[width=5.5cm]{Psir_WS_Bc_He4_4000_1s1p.eps}
\hspace{2ex}
\includegraphics[width=5.5cm]{Psir_WS_Bc_He4_6000_1s1p.eps}
\caption{Coordinate-space wave functions for the 1s and 1p states of the $B_c$-$^{4}$He system
without the Coulomb potentials for
different values of $\Lambda$ calculated using the Fourier transform of
the fitted Woods-Saxon form
potential.}
\label{psirwsbche}
\end{figure}

\begin{figure}[htb!]%
\vspace{6ex}
\includegraphics[width=5.5cm]{Psir_WS_Bc_C12_2000_1s1p.eps}
\hspace{2ex}
\includegraphics[width=5.5cm]{Psir_WS_Bc_C12_4000_1s1p.eps}
\hspace{2ex}
\includegraphics[width=5.5cm]{Psir_WS_Bc_C12_6000_1s1p.eps}
\caption{Coordinate-space wave functions for the 1s and 1p states of the $B_c$-$^{12}$C system
without the Coulomb potentials for
different values of $\Lambda$ calculated using the Fourier transform of
the fitted Woods-Saxon form potential.}
\label{psirwsbcc}
\end{figure}

\begin{figure}[htb!]%
\vspace{6ex}
\includegraphics[width=5.5cm]{Psir_BWS_Bc_He4_2000_1s1p.eps}
\hspace{2ex}
\includegraphics[width=5.5cm]{Psir_BWS_Bc_He4_4000_1s1p.eps}
\hspace{2ex}
\includegraphics[width=5.5cm]{Psir_BWS_Bc_He4_6000_1s1p.eps}
\caption{Coordinate-space wave functions of the 1s and 1p states of the $B_c$-$^{4}$He
system without the Coulomb potentials for
different values of $\Lambda$ calculated using the spherical Bessel transform of
the fitted Woods-Saxon form potential.}
\label{psirbwsbche}
\end{figure}

\begin{figure}[htb!]%
\vspace{6ex}
\includegraphics[width=5.5cm]{Psir_BWS_Bc_C12_2000_1s1p.eps}
\hspace{2ex}
\includegraphics[width=5.5cm]{Psir_BWS_Bc_C12_4000_1s1p.eps}
\hspace{2ex}
\includegraphics[width=5.5cm]{Psir_BWS_Bc_C12_6000_1s1p.eps}
\caption{Coordinate-space wave functions for the 1s and 1p states of the $B_c$-$^{12}$C system
without the Coulomb potentials for
different values of $\Lambda$ calculated using the spherical Bessel transform of
the fitted Woods-Saxon form potential.}
\label{psirbwsbcc}
\end{figure}

\begin{figure}[htb!]%
\vspace{6ex}
\includegraphics[width=5.5cm]{Psir_Bc_AttCoul_C12_2000_1s1p.eps}
\hspace{2ex}
\includegraphics[width=5.5cm]{Psir_Bc_AttCoul_C12_4000_1s1p.eps}
\hspace{2ex}
\includegraphics[width=5.5cm]{Psir_Bc_AttCoul_C12_6000_1s1p.eps}
\\
\vspace{8ex}
\includegraphics[width=5.5cm]{Psir_Bc_RepCoul_C12_2000_1s1p.eps}
\hspace{2ex}
\includegraphics[width=5.5cm]{Psir_Bc_RepCoul_C12_4000_1s1p.eps}
\hspace{2ex}
\includegraphics[width=5.5cm]{Psir_Bc_RepCoul_C12_6000_1s1p.eps}
\caption{Coordinate-space wave functions for the 1s and 1p states of the $B^{\pm}_c$-$^{12}$C
systems with the Coulomb potentials for
different values of $\Lambda$ calculated using the direct Bessel transform.}%
\label{coulwfn}%
\end{figure}

The three methods discussed above and the cutoff dependence in the nuclear potential
give slightly different results for the bound-state energies
and wave functions of the meson-nucleus systems considered.
However, we emphasize that these
differences do not change the conclusion that
the studied meson-nucleus systems should form bound states
based on the strong potentials.

We compare in Fig.~\ref{compwvf} the wave functions of the 1s state of the
$\Upsilon$-$^{12}$C system for different cutoff values and methods.
Although the bound-state energies of the 1s energy level varies slightly according
to the cutoff values or the method used, the wave functions
obtained show very similar shapes and amplitudes.

\begin{figure}[htb!]%
\vspace{6ex}
\centering
\includegraphics[width=6.0cm]{Lambda_Comp_Psir_Y_C12_1s.eps}
\hspace{8ex}
\includegraphics[width=6.0cm]{Method_Comp_Psir_Y_C12_2000_1s.eps}
\caption{
1s state wave functions of the $\Upsilon$-$^{12}$C system for different cutoff values (left)
and momentum-space transformation methods (right).
\label{compwvf}
}
\end{figure}

\section[\appendixname~\thesection]{Numerical precision}
\label{ApxB}

In Table~\ref{errtbl} we compare the bound-state energies of the 1s state of the 
$\Upsilon$-$^{12}$C system for different maximum error acceptances in
the convergence of the results and number of grid points used in the numerical
integrations. We see that the results are numerically consistent within
each method. (See Table~\ref{tblups} for comparison.)

\begin{table}[htb!]
\label{errtbl}
\caption{The 1s $^{12}_{\Upsilon}\text{C}$
bound state energies, obtained by the direct Bessel
  [Eq.~(\ref{bsstr})] and Woods-Saxon Fourier 
  [Eqs.~(\ref{wsfr}) and~(\ref{frdcmp})] transform methods
for $\Lambda$ = 2000 MeV and different values of numerical error acceptance
(indicated by ''Error'' in the Table) and number of
integration points. (See Table~\ref{tblups} for comparison.)
}
\begin{center}
\begin{tabular}{ll|r|r}
  \hline \hline
  & & \multicolumn{2}{c}{Bound state energies (GeV)} \\
  \hline
& & \multicolumn{2}{c}{Direct Bessel transform} \\
\hline
& Error & NP = 100 & NP = 200 \\
\hline
& $10^{-5}$ & -0.013222572875961625 & -0.013222565247691165 \\
& $10^{-8}$ & -0.013222564819422189 & -0.013222564814378224 \\
& $10^{-16}$ & -0.013222564814095783 & -0.013222564814095783 \\
\hline
& & \multicolumn{2}{c}{Woods-Saxon Fourier transform} \\
\hline
& Error & NP = 100 & NP = 200 \\
\hline
& $10^{-5}$ & -0.010506281849386845 & -0.010506341768125260 \\
& $10^{-8}$ & -0.010506261966073183 & -0.010506321862665757 \\
& $10^{-16}$ & -0.010506261926916061 & -0.010506321823470444 \\
\hline
\end{tabular}
\end{center}
\end{table}


\end{document}